\documentclass{elsart1p}
\usepackage{graphicx}
\usepackage{amssymb}
\begin{document}
\begin{frontmatter}
%
%
%
%
%
\title{Entropy scaling from chaotically produced particles in p-p collisions at LHC energies}
%
%

\author{Supriya Das\corauthref{cor1}}, 
\author{Sanjay K. Ghosh, Sibaji Raha and Rajarshi Ray}
\corauth[cor1]{supriya@bosemain.boseinst.ac.in}
\vspace {0.2cm}
\address{Department of Physics and Center for Astroparticle Physics \& Space 
Science, Bose Institute, Kolkata, India} 

\begin{abstract}
Scaling of information entropy obtained from chaotically produced particles 
in p-p collisions, has been shown to be valid up to the highest available
collision energy at LHC. Results from Monte Carlo simulation model PYTHIA 6.135
have also been compared. Based on the two component model and collision
energy dependence of the chaoticity, charged particle multiplicities at
proposed higher collision energies have been predicted.
\end{abstract}
\end{frontmatter}

\section{Introduction}
\label{int}
Multiparticle production in high energy hadron-hadron, hadron-nucleus and
nucleus-nucleus collisions has been in the center of interest in the field of
experimental high energy physics programs for the last
four decades. But in spite of this rigorous exercises for such a long period,
no satisfactory as well as consistent theory has emerged to explain the
results from these studies. Although the quantum chromodynamics (QCD) has been
accepted as the proper theory for the strong interaction envisaged in these
collisions, it is still not known how to treat the soft, thus non-perturbative
processes. Moreover, the investigations of the equation of state of the matter
produced in these collisions has become a matter of increased interest because
of its connection with the studies on Quark Gluon Plasma \cite{plumer1,plumer2}. 
Trace of deconfinement in hadronic collisions at TeV energies have already been 
reported earlier \cite{porile}. Recently very high multiplicities and other new 
features have become the central matter of discussions after the data from LHC 
have come out \cite{alice, cms}. Considering these facts, one could say that the 
experimental data in this particular field are
more important than any other. Analyses of these data using statistical
moments \cite{ua51,ua52} and scaling laws provide new ideas to interpret the
results. At the beginning of this exercise, Bjorken scaling was accepted to be
the tool to explain the picture of parton degrees of freedom.  The collision
energy dependence of the multiplicity of the produced particles was nicely
explained by KNO scaling \cite{kno} up to the ISR energies. But this scaling
law broke down \cite {ua51, ua52} once the collision energy
increased. A new quantity, entropy \cite{simak}, was proposed to revive the
scaling of multiplicity distributions in high energy hadronic collisions at
the collider energies. However, the new scaling holds good for the entire range of
energy starting from ISR ($\sqrt{s}$ = 19 GeV) up to the highest available
collision energy at SPS ($\sqrt{s}$ = 900 GeV) only if the entropy is calculated
from the chaotically produced particles \cite{raha}.  
In this paper we will discuss the results from this scaling law for the 
multiplicity distributions for p-p collisions at the available collision energies 
at LHC.

\section{Formalism}
\label{form}
The scaling variable, {\it information entropy}, has been calculated within the
context of the two component model \cite{fowler} for multiplicity
distribution. According to this model the emission of particles occurs from a
convolution of a chaotic source with $k$ = 1 or 2 and a coherent one mode
source. The entropy for multiplicities in symmetric pseudorapidity intervals 
$|\eta|\leq \eta_c$ is given by

\begin{eqnarray}
\nonumber S(\eta_c,\sqrt{s})
\nonumber &=& (n_{ch}(\eta_c,\sqrt{s}) +1)ln(n_{ch}(\eta_c,\sqrt{s}) +1) \\
&-&  n_{ch}(\eta_c,\sqrt{s}) ln n_{ch}(\eta_c,\sqrt{s})
\end{eqnarray}

where $n_{ch}$ is the chaotic fraction of the average multiplicity. This is
obtained from the total multiplicity by

\begin{equation}
n_{ch} = \tilde{P} n 
\end{equation}

where

\begin{equation}
\tilde{P} = [k \{C_2 - (1+1/\langle n \rangle) \}]^{1/2}
\end{equation}

and $C_2 = \langle n^2 \rangle/\langle n \rangle^2$, the second moment of
multiplicity distribution and $\langle n \rangle$ is the mean of the
distribution. Then we plot $S/\eta_{max}$ as a function of
$\xi = \eta_c/\eta_{max}$ where

\begin{equation}
\eta_{max}=ln[(\sqrt{s} - 2m_n)/m_\pi]
\end{equation}

\section{Results}
\label{results}

\begin{center}
\begin{figure}[htbp]
\includegraphics[scale=0.5]{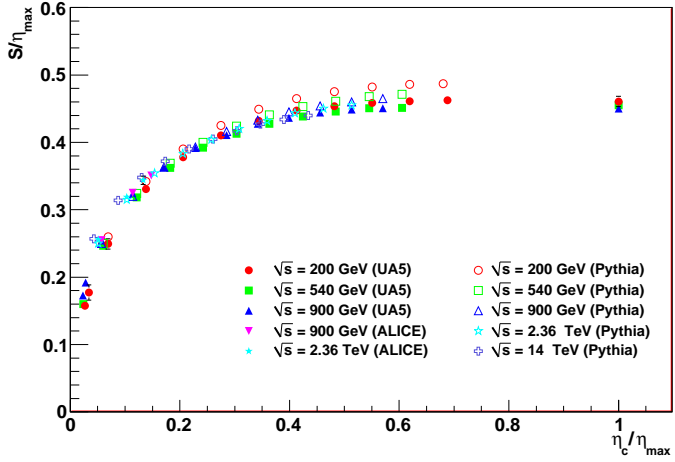}%
\caption{\label{fig:enscaling}Entropy scaling from two component model. Results from 
data and simulation have been shown with ``solid'' and ``open'' symbols respectively.}
\end{figure}
\end{center}

In Fig.~\ref{fig:enscaling} the $S/\eta_{max}$ has been plotted as a function 
of $\eta_c/\eta_{max}$ for different
collision energies starting from SPS \cite{ua51,ua52} up to the highest
available energy at LHC \cite{alicedata}. This figure shows that the results from of
all different collision energies fall on a single curve within the estimated
statistical errors. We have also plotted the results obtained from Monte Carlo 
simulation program PYTHIA 8.135 \cite{pythia} for the same energies as well as the 
projected highest collision energy at LHC ($\sqrt{s}$ = 14 TeV). PYTHIA shows 
slight differences at higher pseudorapidities.

\begin{center}
\begin{figure}[htbp]
\includegraphics[scale=0.5]{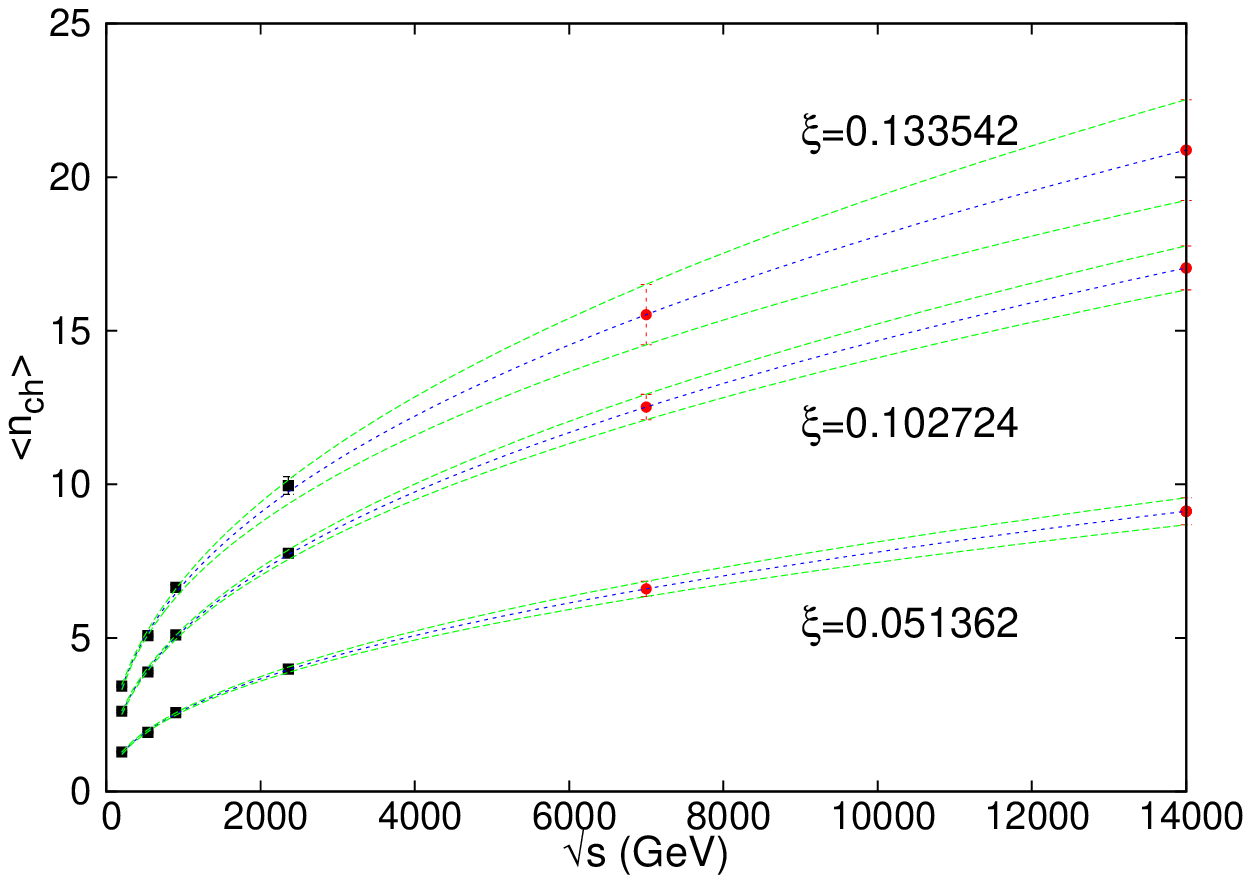}%
\includegraphics[scale=0.5]{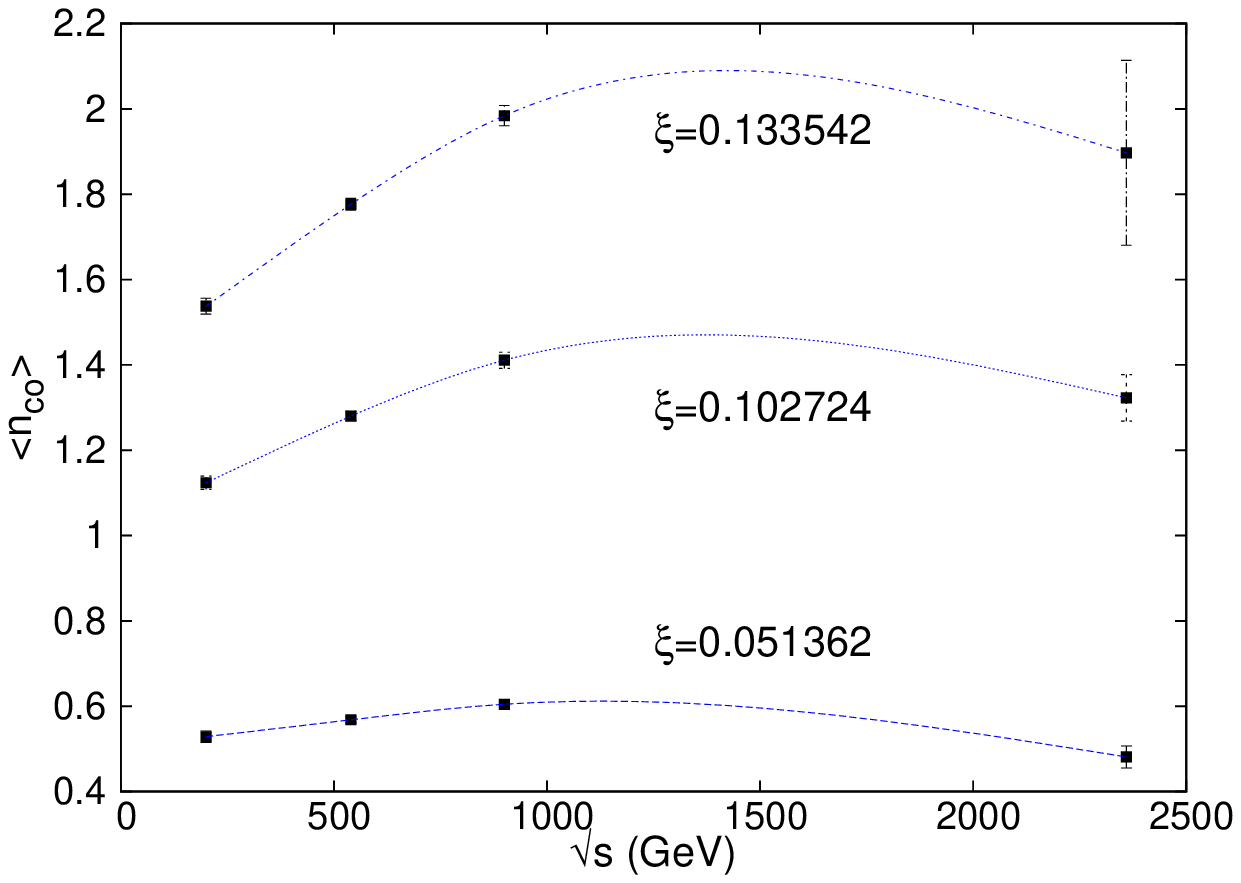}%
\caption{\label{fig:nch}Mean chaotic multiplicity, $n_{ch}$
  (left) and mean coherent multiplicity $n_{co}$(right), 
 as a function of collision energy. The black (square) and red (circle) points 
 correspond to data and predicted points respectively and the green bands show the 
 limits of statistical errors. The lines in the right plot are to
 guide the eye. }
\end{figure}
\end{center}

After the new scaling has been established, we have plotted the mean
multiplicity of chaotically produced particles ($n_{ch}$) as a
function of collision energy for different values of $\xi$
(Fig.~\ref{fig:nch}). Here we have plotted this quantity for three different
$\xi$ values for which data are available from ALICE experiment. Fitting
this distribution with a power law, one obtains the $n_{ch}$ at higher collision
energies that will be obtained at LHC in near
future. We have also plotted the other fraction of multiplicity {\it i.e.} the 
multiplicity of coherently produced particles as a function of collision
energy in the similar way. This quantity shows a non monotonous behaviour as
the energy increases beyond 1 TeV. This tells us that beyond this point the 
chaotic fraction of multiplicity predominates. At collision energy 14 TeV the 
chaoticity parameter $\tilde{P}$ reaches very close to unity.

We understand that the more important quantity from experimental point of view
is the average of total multiplicity as that could be measured directly in the
experiments. Keeping this in mind we have finally plotted that quantity again
as a function of energy for available collision energies and fitted that
distribution again with a power law. From the fitting we predict the average 
multiplicity in p-p collisions at 7 TeV and 14 TeV collision energies and most 
central pseudorapidity bin to be 6.678 $\pm$ 0.242 and 8.658 $\pm$
0.404 respectively.

\begin{center}
\begin{figure}[htbp]
\includegraphics[scale=0.5]{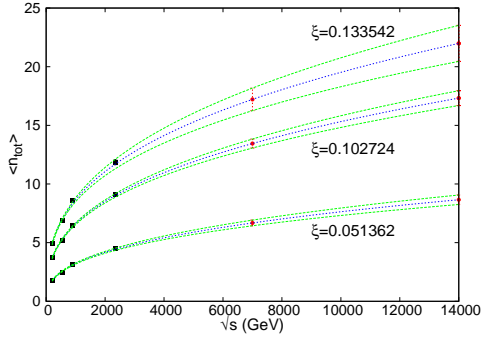}%
\caption{\label{fig:ntot}Mean total multiplicity, $n_{tot}$,
  as a function of collision energy. The black (square) and red (circle) points correspond to
 data and predicted points respectively. }
\end{figure}
\end{center}

\vspace {-0.7cm}

\section{Summary}

In summary we have applied the two source model to extract the multiplicity of the 
chaotically produced charged particles at the LHC energies. The information entropy 
obtained from this scales nicely up to the highest available collision energy at
LHC. The chaoticity parameter obtained at LHC energies is close to unity signifying
the possibility of collective phenomena in hadronic collisions at these energies. We 
have gone further to predict the multiplicities for the proposed higher energy
collisions at LHC.  

\section{Acknowledgements}  

This work was partly supported by Department of Science and Technology,
Govt. of India under IRHPA scheme.

\end{document}